\documentclass[runningheads]{llncs}
\usepackage{booktabs,tabularx}
\usepackage{makecell}
\usepackage{amsmath,amsfonts,amssymb}
\usepackage{graphicx}
\usepackage[usenames,dvipsnames]{xcolor}
\usepackage{multirow}
\usepackage[normalem]{ulem}
\usepackage{float}
\usepackage{capt-of}

\newcommand{\etal}{\emph{et al.} } 
\newcommand{\ie}{\emph{i.e.}, }
\newcommand{\eg}{\emph{e.g.}, }

\begin{document}
\title{Gender Patterns of Human Mobility in Colombia: Reexamining Ravenstein's Laws of Migration }
\titlerunning{Effects of gender on urban mobility}
%
\author{Mariana Macedo\inst{1}\orcidID{0000-0002-7071-379X} \and
Laura Lotero\inst{2}\orcidID{0000-0002-6537-3276} \and
Alessio Cardillo\inst{3,4,5}\orcidID{0000-0003-4811-9978} \and
Hugo Barbosa\inst{1}\orcidID{0000-0002-3927-969X} \and
Ronaldo Menezes\inst{1}\orcidID{0000-0002-6479-6429}}
\authorrunning{M. Macedo et al.}
%
\institute{BioComplex Lab, Department of Computer Science, University of Exeter, UK \and
Faculty of Industrial Engineering, Universidad Pontificia Bolivariana, Colombia \and
Department of Engineering Mathematics, University of Bristol, UK \and
Dept. of Computer Science and Mathematics, University Rovira i Virgili, Spain \and
GOTHAM Lab, Institute for Biocomputation and Physics of Complex Systems, University of Zaragoza, Spain %
}
\maketitle              
\begin{abstract}
Public stakeholders implement several policies and regulations to tackle gender gaps, fostering the change in the cultural constructs associated with gender. One way to quantify if such changes elicit gender equality is by studying mobility. In this work, we study the daily mobility patterns of women and men occurring in Medell\'in (Colombia) in two years: 2005 and 2017. Specifically, we focus on the spatiotemporal differences in the travels and find that {\em purpose of travel} and {\em occupation} characterise each gender differently. We show that women tend to make shorter trips, corroborating Ravenstein's Laws of Migration. Our results indicate that urban mobility in Colombia seems to behave in agreement with the "archetypal" case studied by Ravenstein.
%
\end{abstract}
%
%
%

%
%

\section{Introduction}


Our daily lives are shaped by the convolution of a broad range of individual and social-level demands (\eg eat, sleep, work, pay bills), and mobility is essential for their fulfilment. Hence, the betterment of our lives passes through the study of how people move \cite{BARBOSA20181,toch2019analyzing}. Some models on mobility assume that all travellers are -- more or less -- the same, disregarding the wealth of attributes discriminating one social group from another. Conversely, other studies demonstrate that social demographic attributes like socio-economic status do play a role in mobility \cite{lotero2016several}. Amidst the plethora of attributes available, gender is a key one because men and women can emerge alternative behaviours \cite{maharani2019analyzing}. Recently, the gap existing between men and women has become the focus of many studies (\eg on urban mobility and academic performance) \cite{evertsen2019gender,gauvin2019gender,o2018gender,thynell2016quest}. Despite the efforts made to improve gender equality \cite{care}, both gender's routines remain being affected differently \cite{gauvin2019gender}. Understanding such differences is crucial to build a better ``environment'' for everyone \cite{evertsen2019gender,levy2013transport,turner2012urban}, and design interventions on transportation aimed at reducing gender gaps to offer the same mobility opportunities \cite{milan2017lifting,thynell2016quest}.

In 1885, Ravenstein published a paper entitled ``\emph{The Laws of Migration}'' \cite{ravenstein1885laws} where he highlighted differences between women's and men's mobility arguing that, on average, women migrate more than men. According to Ravenstein, women were more likely to visit areas nearby their ``homes,'' mainly with the purpose of seeking for jobs. Accordingly, women mainly migrated to residential and job rich areas (\eg industrial). However, society has undergone deep transformation since Ravenstein's study.


In this work, we study the gender-based spatiotemporal differences in urban mobility taking place within the Medell\'in's metropolitan area (known as Aburr\'a Valley) in Colombia in two distinct years, 2005 and 2017. Using an approach similar to the one used in \cite{lotero2016several,lotero2016rich}, we find that despite more than one century has passed since Ravenstein's study, women still make shorter trips -- remaining closer to their ``homes,'' -- and that employment continues to play a considerable role in mobility.  Furthermore, our results are in agreement with those of a recent study based on Chilean mobile phone data by Gauvin \etal \cite{gauvin2019gender} which found, among other things, that men tend to visit more diverse places than women.

%
%
 
\section{Dataset}

\begin{figure}[h!]
\centering
\includegraphics[width=0.9\textwidth]{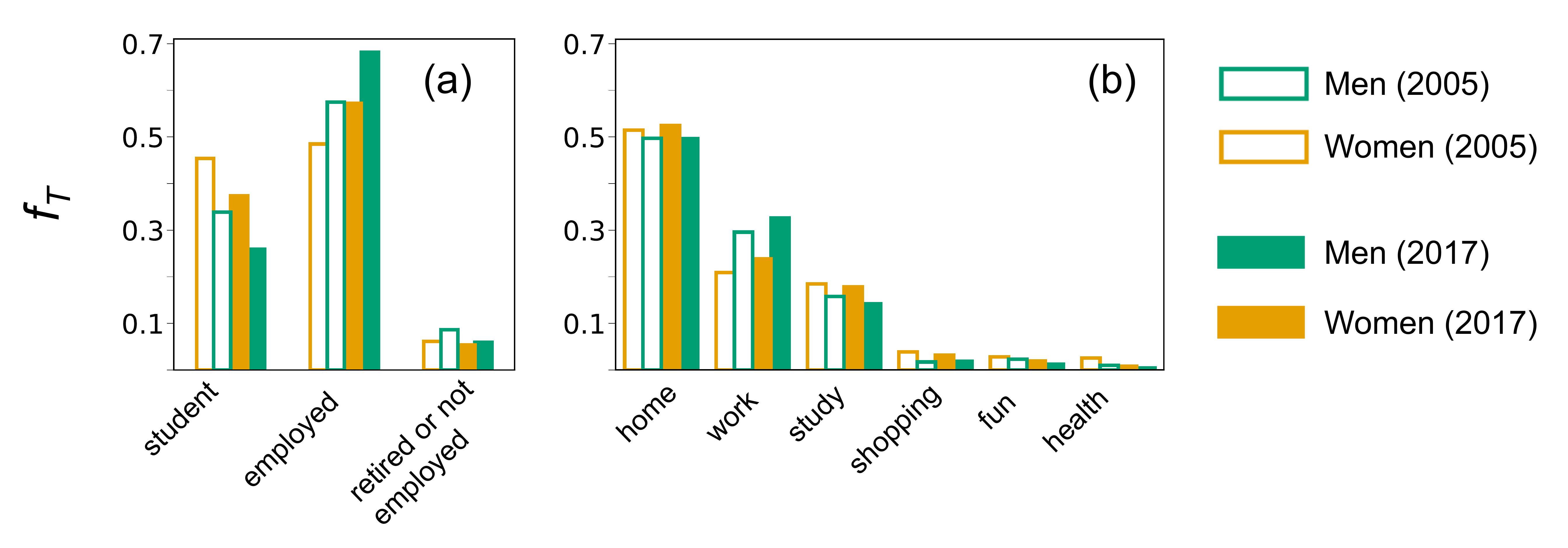}
\caption{Fraction of travels made ($f_T$) by each gender grouped either by occupation (panel \textbf{(a)}), or travel's purpose (panel \textbf{(b)}). The empty bars accounts for data coming from 2005's survey, whereas the filled bars accounts the 2017 case, instead.}
\label{fig:occupation_purpose}
\end{figure}

We consider the data collected by two surveys on urban mobility made within the Aburr\'a Valley's metropolitan area surrounding the city of Medell\'in (Colombia). Each survey accounts for a distinct year, namely 2005 and 2017. Each interviewed householder is asked about the travels she/he had the day before the interview, providing with the origin and destination zones, the departure and arrival times, the transportation mode used, and the purpose of each travel. In addition, householders are characterised by their socio-demographic characteristics (age, gender, and occupation) which define their \emph{socioeconomic status} (SES). Each \emph{travel} is divided into one or more \emph{trips} each corresponding to a displacement made with a specific transportation mode. For example, if one traveller went from zone $A$ to zone $B$ walking during the first part and then took a bus, the corresponding travel is made of two trips. Table \ref{tab:dataset} summarises the main features of each survey such as: the number of travellers ($N_P$), the number of travels ($N_T$), and their composition in terms of gender.\newline
%
%
%
%
\begin{table}[ht]
  \centering
    \caption{Datasets' main properties. For each year, we report the number of zones $N_Z$, the total area covered, the number of travellers $N_P$, the fraction of men (women) travellers $f^M$ ($f^W$), the number of travels $N_T$, and the fraction of travels made by men (women) $f_{T}^M$ ($f_T^W$).}
    \label{tab:dataset}
      \tabcolsep=0.24cm
       \resizebox{\textwidth}{!}{
        \begin{tabular}{l c c c c c c c c}
            \toprule
            \textbf{Year} & $N_Z$ & \makecell{Total surface \\ ($km^2$)} & $N_P$ & $f^M$ & $f^W$ & $N_T$ &
            $f_{T}^M$ & $f_{T}^W$ \\
            \midrule
            \textbf{2005} & 403 & 1,043 & 55,681 & 0.5143 & 0.4854 & 126,164 & 0.5163 & 0.4833 \\
            \textbf{2017} & 521 & 1,174 & 30,107 & 0.5418 & 0.4582 & 64,837 & 0.5434 & 0.4566 \\ 
            \bottomrule
        \end{tabular}
     }
\end{table}
\indent Leveraging the meta information available, we can divide the dataset into several subsets based, \eg on individual occupation or travel's purpose. In Fig.~\ref{fig:occupation_purpose}, we display the percentage of travels made by each gender grouped either according to the traveller's occupation (panel a), or to travel's purpose (panel b). It is worth noting that the categories displayed in Fig.~\ref{fig:occupation_purpose} correspond to 100\% of 2005's data, and to 90.36\% for occupations and to 92.10\% for purpose, in 2017, instead. In 2017, the dataset contains other occupations and trips' purpose such as \texttt{housewives} and \texttt{give a ride to someone} which are absent in 2005's dataset. According to Fig.~\ref{fig:occupation_purpose}(a), the majority of our samples is made of students and workers, with more women in the student's group and men in the worker's group (the worker's group showing a rise between 2005 and 2017).
%
%
%
The predominance of \texttt{students} and \texttt{employed} travellers reverberates on the frequency of travel's purposes (Fig.~\ref{fig:occupation_purpose} (b)), with \texttt{work} and \texttt{study} being the second and third most frequent classes, preserving the imbalance between genders in both years. Other travel's purposes seem quite underrepresented, probably due to their less periodic nature. However, on average, the fraction of travels made by women due to \texttt{other} purposes is higher than the men counterpart. Despite the gender imbalance across occupations and purposes of travels, we show evidences in the following sections that gender can play a role in mobility not because of the abundance of travellers in a class but, rather, because women and men behave differently in our sample (\ie have distinct routines).

\section{Network description}

Travels occur between zones and the Aburr\'a Valley is divided into $N_Z$ zones. However, there are differences between the zones partitioning in 2005 and 2017, with 2017 displaying a more granular structure due to the growth of the city and its metropolitan area. Since the vast majority of 2017's zones are subsets of 2005 ones, to ensure compatibility between the results, we consider the 2005 zone structure on both surveys. Nevertheless, it is worth noting that the use of 2005 partitioning on 2017 data does not alter significantly the overall behaviour of the distributions of travel's distance, number of locations visited, and number of transport modes used per travel.\newline
%
%
%
%
\begin{figure}[b!]
\centering
\includegraphics[width=0.8\textwidth]{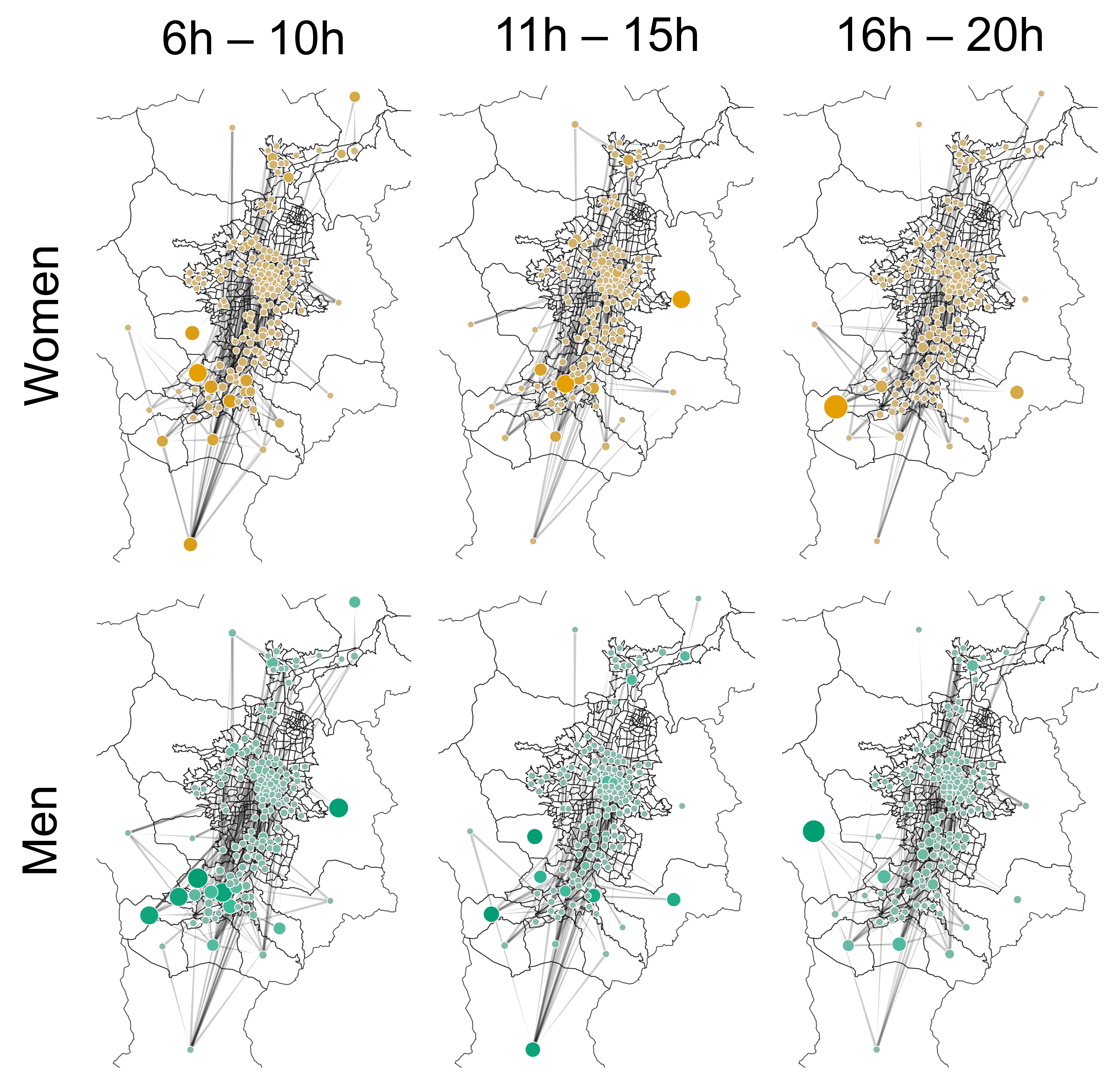}
\caption{Network representation of gender mobility \emph{flows} (\ie number of travels) occurring during the morning (left column), midday (central column), and afternoon (right column). The nodes' size accounts for the in-strength of a zone (\ie the sum of the weights of all edges entering in a zone), while the edge thickness and colour accounts for the number of travels made between two zones. Data refers to the 2017's survey.}
\label{fig:network}
\end{figure}
\indent The mobility data can be mapped into a weighted spatial network where zones are the nodes, and travels between zones represent the edges \cite{Barthelemy2011}. Each edge can be associated with two attributes (weights): the distance between the origin and destination zones, and the number of travels made between those two zones. For each year, we consider initially three different networks: one accounting for the whole mobility, and other two accounting for men and women travels only. In Fig.~\ref{fig:network}, we present an example of such networks where we display the flow of travels made during certain hours of the day.\newline
%
%
%
\indent In 69.96\% of the zones, the men's flow is higher than its women counterpart. The set of departure and arrival zones visited by men and women are statistically different according to the Kolmogorov Smirnov (KS) test with a confidence level of 95\% and p-values of $6.07 \cdot 10^{-14}$ and $7.87 \cdot 10^{-14}$, respectively. If we differentiate travels according to their purpose, the differences between the arrival zones of men and women are higher for employed people and for \texttt{work} purpose (KS with p-values of $1.53 \cdot 10^{-90}$ and $8.55 \cdot 10^{-23}$, respectively). The differences between the departure zones are even higher when we consider the \texttt{home} travel's purpose (KS with p-value of $2.32 \cdot 10^{-12}$).


\section{Spatial characterisation of travels}

\subsection{Analysis of travel's distance}

\begin{figure}[hbt!]
\centering
\begin{minipage}[c]{0.49\textwidth} 
\includegraphics[width=0.95\textwidth]{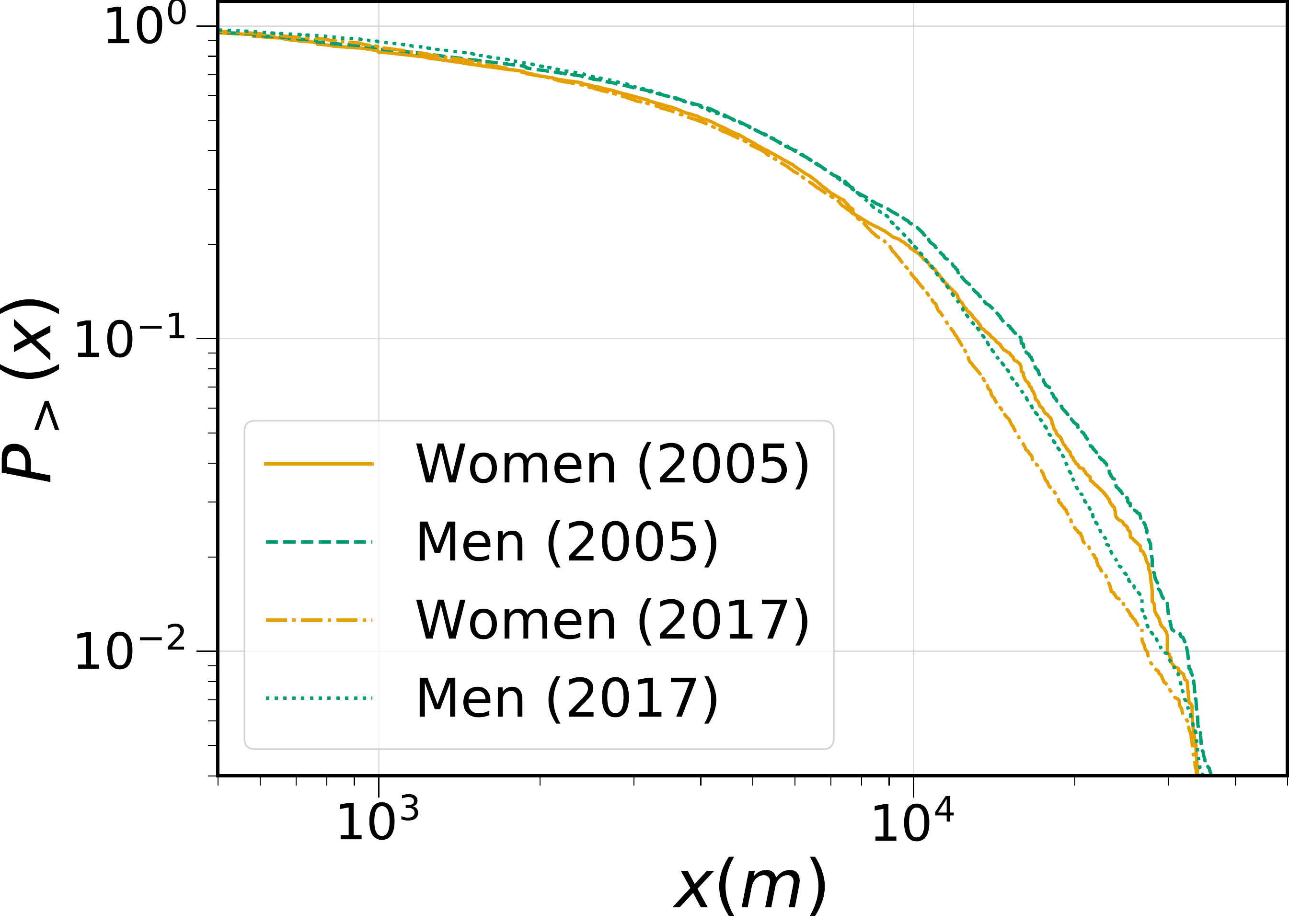}
\captionof{figure}{Complementary Cumulative distribution function, $P_> (x)$, of the probability of making a travel with a distance between the origin and destination zones equal to $x$.}
\label{fig:shortlongtrips}
\end{minipage}
\hfill
\begin{minipage}[c]{0.47\textwidth} 
\captionof{table}{Average values of travel's distance, $\langle x \rangle$ (m), for travels made by men (M) and women (W) in each year.}
    \label{tab:avg_dist}
      \tabcolsep=0.25cm
       \resizebox{\textwidth}{!}{
        \begin{tabular}{l c c }
            \toprule
            \textbf{Year} & gender & $\langle x \rangle$ (m) \\
            \midrule
            \multirow{ 2}{*}{\textbf{2005}} & M & 6711.42 \\
             & W & 6031.75 \\
            \midrule
            \multirow{ 2}{*}{\textbf{2017}} & M & 6319.32 \\ 
             & W & 5542.41 \\
            \bottomrule
        \end{tabular}
     }
\end{minipage} 
\end{figure}

One of Ravenstein migration's laws postulates that women tend to make short travels. Here, we check whether this is the case also for Medell\'in's urban area. To this aim, we compute the length, $x$, of each travel as the distance between the centroids of its origin and destination zones. We considered only distances greater than 100 meters, to avoid underestimated displacements. Moreover, we do not account for travels made within the same zone because we cannot estimate their displacements. After that, we compute the complementary cumulative distribution function (CCDF) quantifying the probability that a travel has length longer than $x$, $P_> (x)$ (Fig.~\ref{fig:shortlongtrips}). Looking at the CCDF, we do not observe any remarkable differences between them (confirmed also by a KS test). However, the averages of the travel's distances, $\langle x \rangle$, displayed in Tab.~\ref{tab:avg_dist}, indicate that -- on average -- men tend to perform longer travels than women.

%
%
%
We validate such claim with a $t$-test which returns p-values of $1.19 \cdot 10^{-8}$ for 2005 and $3.48 \cdot 10^{-11}$ for 2017, instead. When we consider also travels of less than 100 meters, the differences between the genders are amplified.

The gender asymmetry between short and long range travels is preserved also when we discriminate them according to either travellers' occupation or travel's purpose (results not shown). If we further divide mobility according to departure time, we find that men are more prone to make short travels between 23h and 4h. Such difference might be related with the fact that women in Colombia feel more insecure to move at late night and early morning \cite{heinrichs2014public}.

\subsection{Spatial coverage}

Another relevant aspect is how travels sprawl in space. Overall, we observe that men tend to visit more distinct (unique) zones than women. Such differences can be quantified by computing the Shannon entropy, $S^X$, of the sequence of zones visited by each person with gender $X = \{ M, W\}$ \cite{shannon1948mathematical} which, up to a multiplicative factor, reads as:
\begin{equation}
\label{eq:entropy}
S^X = - \sum_{i=1}^{N_Z} p^X(i) \; \log_{2} p^X(i) \,.
\end{equation}
%
%
%
%
%
Where $p^X(i)$ is the probability that zone $i$ is visited by a travel made by travellers with gender $X$, which is:
\begin{equation}
\label{eq:probability}
p^X (i) = \dfrac{{N_T}^X (i)}{{N_T}^X} \,,
\end{equation}
where ${N_T}^X$ is the total number of travels made by gender $X$, and ${N_T}^X (i)$ is the number of those that visit zone $i$. %
For each year, we compute the entropy of all travels made regardless of traveller's gender ($S$), of travels made by men ($S^M$), and by women ($S^W$). Then, we compute the entropy difference $\Delta S^X = S^X - S$ and study its sign. Table \ref{tab:entropy} displays the values of $S$, $S^X$, and $\Delta S^X$ for both genders and years.
%
%
%
%
%
\begin{table}[h!]
  \centering
  %
  %
  %
  %
  \caption{Entropy of travels made by all travellers, $S$, or by men, $S^M$ and women, $S^W$, only. We display also the differences between the entropy of gender $X$ and of the whole population, $\Delta S^X$. Each row accounts for a different survey (year).} 
  \tabcolsep=0.2cm
\resizebox{.8\textwidth}{!}{
  \begin{tabular}{c c c c c c}
  \toprule
\textbf{Year} & $S$ & $S^M$ & $S^W$ & $\Delta S^M$ & $\Delta S^W$ \\
     \midrule 
\textbf{2005} & 7.7761 & 7.7657 & 7.7736 & -0.0024 & -0.0103 \\
\textbf{2017} & 6.7359 & 6.7444 & 6.7209 & 0.0085 & -0.0149 \\
\bottomrule
  \end{tabular}
  }
  \label{tab:entropy}
\end{table}

Given the values displayed in Tab.~\ref{tab:entropy}, in 2017, men displacements appears to be slightly more ``explorative'' (\ie more entropic) than women's ones. In fact, women tend to return 2-4\% more frequently to zones closer to their origins. The women's tendency to return to their origin zones is stronger in 2017 than in 2005. In agreement with such trend, if we account also for the travels made within the same zone, we observe 5\% more self loops in the women's network than in the men's one. In 2005, instead, men tend to be slightly less entropic than women, albeit such difference is not very high. The analysis of entropy alone is not entirely conclusive but, based on other evidences, we argue that in our case study men's mobility seems to be more explorative than women's one.

\subsection{Transportation multimodality}

\begin{figure}[t!]
\centering
\includegraphics[width=0.95\textwidth]{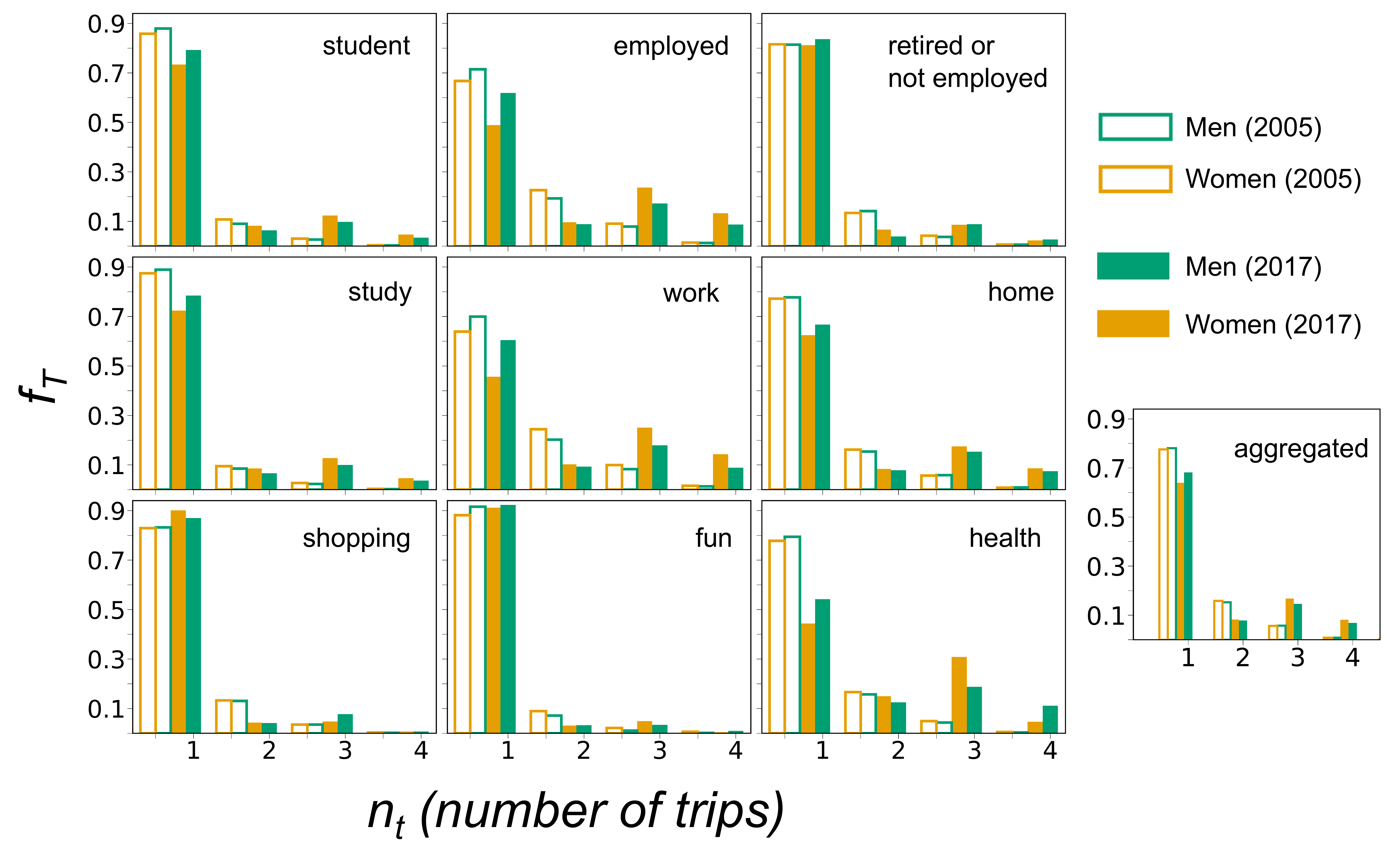}
\caption{Histograms of the fraction of travels made ($f_T$) by $n_t$ trips for several occupations and travel's purposes. The rightmost histogram (\ie \texttt{aggregated}) accounts for the the data aggregated altogether. Empty and filled bars refer to data from the 2005's and 2017's survey, respectively. We use travels made of at most four trips corresponding to the 97.32\% and 99.92\% of the whole dataset for 2005 and 2017, respectively.}
\label{fig:steps}
\end{figure}

There is a difference between men and women in the usage of certain transportation modes. For example, men tend to use the car twice much than women, and men reach \emph{directly} their destination more often than women. However, to reach its destination, one might need to use more than a single transportation mode. Here, we quantify the transportation multimodality of each gender. Fig. \ref{fig:steps} reports the histograms of the fraction of travels made of $n$ trips for occupations and travel's purposes. A quick glance at the histograms reveals that the bulk of travels ($\gtrsim 60\%$) is made by a single trip, regardless of traveller's occupation, purpose, and gender. Another feature highlighted by the histograms is that men are more inclined to reach their destinations using a single transportation mode, whereas women are more likely to use between 3 and 5 transportation modes. For example, in 2017, 69\% of \texttt{employed} men used one transportation mode (mainly, motorcycle), while only 63\% of women did the same (mainly, walking). 

Occupation wise, \texttt{students} and \texttt{retired}/\texttt{unemployed} people reach their destinations mostly directly, whereas \texttt{employed} tend to use more than one transportation mode. Travel's purposes display trends similar to those observed for occupations, with \texttt{shopping}, \texttt{fun}, and -- to some extent -- \texttt{study} appearing more ``direct'', while the others (especially \texttt{work} and \texttt{health}) tend to involve multiple trips. Finally, the comparison between travels made in 2005 and 2017 highlights the presence of ``longer'' travels -- in terms of number of trips, -- in 2017, suggesting that people have become more multimodal in their displacements. Also, 2017 data display higher gender asymmetry both in terms of number of travels (Tab.~\ref{tab:dataset}) and trips. 
The asymmetry is bigger for travels made of three trips regardless of whether we select them according to occupation/purpose or consider their aggregate form.
%
Finally, the 2017 survey contemplates four additional travel's purposes: \texttt{lunch}, \texttt{bureaucratic} \texttt{activities}, \texttt{accompany someone}, and \texttt{give a ride to someone}. In general, men have lunch more at home than women, and women have a higher amount of travels to perform bureaucratic activities and accompany/pick up someone. However, in the next section, we show that there are also temporal differences in the mobility patterns of men and women.

%
%
%

\section{Temporal characterisation of travels}

\begin{figure}[h!]
\centering
\includegraphics[width=0.95\textwidth]{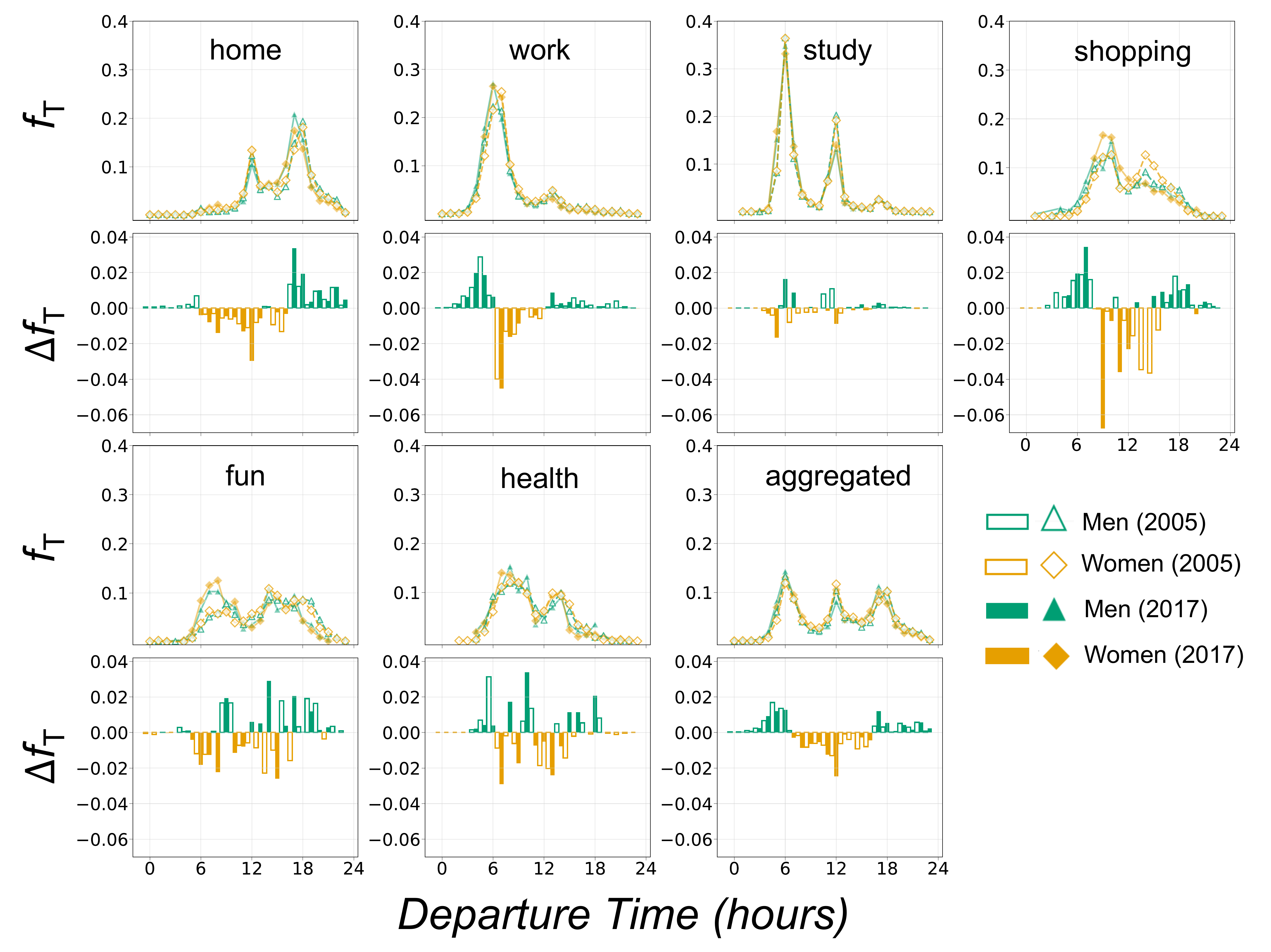}
\caption{Fraction of travels made $f_T$ (line plots), and their gender differences $\Delta \, f_T$ (bar plots), of several travel's purposes and their \texttt{aggregated} case with respect to the departure time (in hours). Empty (filled) symbols refer to data from year 2005 (2017).}
\label{demandsbothyears}
\end{figure}

Travels take place in space but also in time, and their bursty, synchronised, nature is responsible for the emergence of phenomena like traffic jams \cite{heinrichs2014public}. After studying the spatial properties of urban mobility, we focus on the temporal perspective. In particular, we study how women/men travel's purposes reverberate on the ``rythms'' with which travels take place. Fig. \ref{demandsbothyears} portrays the evolution in time of travels made with different purposes. For each purpose, we plot the fraction of travels departing at a given time (line plots), as well as the difference between the fraction of travels made by men and women (bar plots). In this way, we capture both the evolution across time of travels (and eventual longitudinal shifts between genders), as well as any eventual gender-based difference.

\begin{figure}[b!]
\centering
\includegraphics[width=0.95\textwidth]{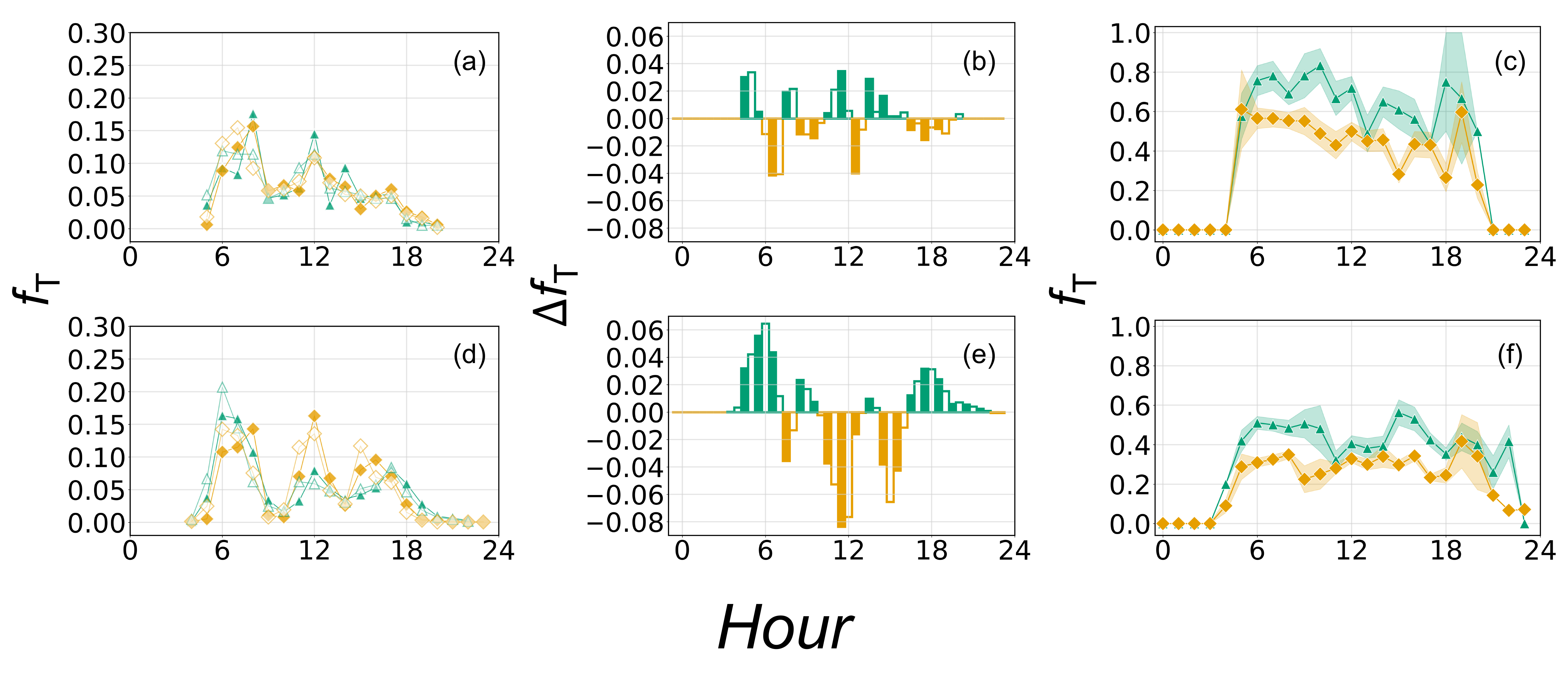}
\caption{
Spatiotemporal features of the travels made in 2017's survey to \texttt{accompany someone} (top row) or \texttt{give a ride to someone} (bottom row). Panels (a) and (d) display the fraction of travels ($f_T$) departing (empty symbols) and arriving (filled symbols) at a given hour. Panels (b) and (e) report the difference ($\Delta f_T$) between the men and women curves displayed in panels (a) and (d). Panels (c) and (f) report the average fraction of travels arriving at a given time averaged over all the available zones (the shaded area denotes the standard deviation over different arrival zones).}
\label{demands2017}
\end{figure}

Eyeballing at Fig.~\ref{demandsbothyears} reveals that each travel purpose has a distinct temporal footprint, with its gender components displaying further differences. Some purposes (\texttt{home}, \texttt{work}, and \texttt{study}) exhibit one -- or more -- clear peak of ``activity'' along the day. For example, returning home occurs mainly around lunchtime and at the end of the afternoon regardless of the gender. Women tend to leave to go to work about one hour later than men, instead. Other less routinely purposes (\texttt{shopping}, \texttt{fun}, and \texttt{health}), instead, display a temporal profile more diluted along the day. Another feature is that for \texttt{home}, \texttt{work}, and \texttt{shopping} purposes, the difference plots highlight the predominance (in proportion) of men's travels during the early morning/late evening, whereas women tend to travel more than men during the middle of the day. Curiously, we observe that the predominance of female travels associated with \texttt{shopping} occurring during the middle part of the day shifts backward from the second half of the day in 2005, to the first half in 2017. The clearly split gender pattern observed along the day is not present for purposes like \texttt{study}, \texttt{fun}, and \texttt{health}, where the middle part of the day is characterised by the lack of prevalence among genders, albeit the \texttt{health} case displays its own unique pattern. \texttt{Study} appears to be the most gender balanced purpose, as denoted by the small values of fractions differences.\newline
\indent The aggregated data display, instead, three peaks located in the morning, midday, and late afternoon exhibiting also a synchronisation of both genders at midday. Men still tend to travel more (in proportion) than women during the morning/evening while women's travels are more prominent during the middle part of the day. Occupation wise, unemployed and retired people distribute their travels along the whole day, with women travelling more often in the interval 12h-18h, and men in the interval 06h-12h, instead. However, the convolution of the temporal curves into the aggregated one smooths out many of the temporal footprints observed, suggesting that gender plays a more prominent role when mobility is studied in terms of people's need/demands. We conclude by noting that we could also make a characterisation in terms of the arrival times. However, such analysis does not highlight any additional feature of mobility.\newline
\indent Finally, we take a step further by presenting a spatiotemporal characterisation of mobility according to two specific travel's purposes recorded exclusively during the 2017's survey: \texttt{accompany someone} and \texttt{give a ride to someone}. Such purposes constitutes benchmarks to highlight gender differences because they presume the use of transportation modes capable to carry people to their destination (like cars and motorcycles). As we have said, men have more access than women to such modes. Moreover, the act of accompany/give a ride to someone implies a social relationship between the carrier and the carried.\newline
%
%
\indent Fig. \ref{demands2017} panels (a) and (d) display the evolution in time of the  fraction of travels associated with the aforementioned purposes, according to the departure and arrival time of each travel, with men and women curves displaying a longitudinal shift only for the \texttt{accompany someone} case. In panels (b) and (e), instead, we report the gender-based differences between the curves displayed in panels a and d. We notice that, in the surroundings of lunchtime women dominate (in proportion) for \texttt{giving a ride} purpose. However, in general, the temporal footprints do not display any feature remarkably different from those appearing in Fig.~\ref{demandsbothyears}. In panels (c) and (f), instead, we display the average of the fraction of all the travels made with a given scope arriving at a given time. The average is computed over the arrival zones. We observe that, for both purposes, men perform more travels (in proportion) than women regardless of the arrival time. \newline
%
%
\indent If we perform a spatio-temporal analysis similar to the panels (c) and (f) of Fig.~\ref{demands2017} for all the travel's purposes displayed in Fig.~\ref{demandsbothyears}, we observe that -- in both years -- women perform (in proportion) more trips from/to \texttt{work} than men. On the other hand, men tend to perform more trips over the day to the same zones on the purposes of travels: \texttt{health} and \texttt{shopping}. The majority of women return \texttt{home} (at same zone) from 14-16h in both years. In general, men tend to move for \texttt{fun} purpose to similar zones at late night, while women tend move for the same purpose from 10-17h, instead. \texttt{Study} remains the most gender neutral travel's purpose.

\section{Conclusions}

In this study, we have analysed the data collected by two surveys on mobility within the Aburr\'a Valley's metropolitan area surrounding the city of Medell\'in (Colombia) to quantify how gender reverberates on urban mobility. Using the network paradigm, we have built networks of travels occurring between distinct zones. By leveraging the wealth of meta-information coming along with the data, we have been able to disentangle not only the contributions associated with each gender, but also how distinct job occupation and demands mold the spatio-temporal features that mobility exhibits.

By analysing the spatial properties of travels, we have found that -- on average -- women prefer to move within or in the proximity of their ``home'' zones, display a more recurrent mobility pattern, and use more transportation modes (mainly walk and bus) than men. Men tend to display a more exploratory behaviour, with a higher number of unique locations visited, more direct travels (mainly made by car or motorbike), and a mobility predominantly characterised by work-related demands (being employed or go to work). The characterisation of the travel's temporal features, instead, unveiled unexpected features induced by gender. The analysis of the percentage of travels made along the day, grouped according to either occupation or purpose, highlights that each class of travels displays a specific temporal pattern, which is different from its aggregated counterpart \cite{lotero2016several}. Such phenomenon constitutes one of the hallmarks of complex systems, which is usually encapsulated by the statement ``\emph{more is different}'' \cite{Anderson393}. Specifically, we observe that -- on average, -- men tend to travel more during the morning/evening, while women, instead, move more during the middle of the day. Such behaviour could be due, on the one hand, to the women's perception of insecurity towards going out during the very early morning or at late night. On the other hand, it could also be due to the higher amount of non-work related travels taking place during the day.
Additionally, the curves of travels associated with the home/work mobility display a temporal shift, with men leaving for work one hour earlier, and women leaving to get back home later than men. Study's purpose appears to be the most gender neutral activity, as denoted by the small differences between the amount of travels made by each gender. Finally, when we analyse the percentage of travels made with a specific purpose averaged over all the departure zones, we notice that men tend to make more travels than women regardless of the departure time. This is in contrast with the same quantities computed without taking into account the zone of departure.

Summing up, we have seen how gender -- in combination with occupation, and demands -- molds urban mobility. Our observations are in agreement with Ravenstein's migration's law \cite{ravenstein1885laws} and recent studies on gender gaps based on mobile phones data \cite{gauvin2019gender}. Nevertheless, the availability of detailed meta-information on travels/travellers allowed us to split mobility into distinct components, enabling a more fine grained analysis of the overall phenomenology observed both in space and time. Of course, a better comprehension of gender effects cannot neglect the influence of other factors like age, education, and socio-economic status. Finally, the availability of further studies concerning different cities/countries, as well as, cultures would improve our understanding the gender effects on mobility by separating them from the spatial environment under scrutiny.


\bibliographystyle{splncs04}
\bibliography{bib}

\end{document}